\documentclass{article}

\usepackage{PRIMEarxiv}

\usepackage[utf8]{inputenc} 
\usepackage[T1]{fontenc}    
\usepackage{hyperref}       
\usepackage{url}            
\usepackage{booktabs}       
\usepackage{amsfonts}       
\usepackage{nicefrac}       
\usepackage{microtype}      
\usepackage{lipsum}
\usepackage{fancyhdr}       
\usepackage{graphicx}       
\usepackage{makecell}
\graphicspath{{media/}}     

\title{Modifications to Image Phase Alignment Super-sampling Produce up to 4.41 times Increased Image Resolution
\thanks{\textit{\underline{Citation}}: 
\textbf{Authors. Title. Pages.... DOI:000000/11111.}} 
}

\author{
  James N. Caron \\
  Quarktet, Silver Spring, Maryland, USA\\
  Caron@Quarktet.com \\
}

\begin{document}


\title{Modifications to Image Phase Alignment Super-sampling Produce up to 4.4 times  \\ Increased Image Resolution} 


\maketitle 

\begin{abstract}  
Image Phase Alignment Super-sampling (ImPASS) is a computational method for combining displaced low-resolution images into a single high-resolution image.  The general steps include measuring the relative displacements, up-sampling, aligning and combining the images, followed by a blind deconvolution.  Previous ImPASS studies have shown that the resulting image resolution can significantly subceed the diffraction limit of the imaging system.  Characteristics that potentially limit the processed image resolution include optical parameters, detector noise, image alignment accuracy, or deconvolution parameters.  In this report, modifications have been made to the algorithm to improve the image alignment accuracy and deconvolution.  Applications of the modified algorithm improved image resolution by a factor up to 1.81.  Compared to the original image resolution, the modified ImPASS achieved a resolution improvement factor up to 4.41 while subceeding the diffraction limit by a factor of 2.57.  This suggests that limitations imposed by the physical properties of the system have not yet been reached, and further improvement of the algorithm is warranted.
\end{abstract}

\section{Introduction}
\begin{figure}[htbp]
\center{\includegraphics[width=24pc]{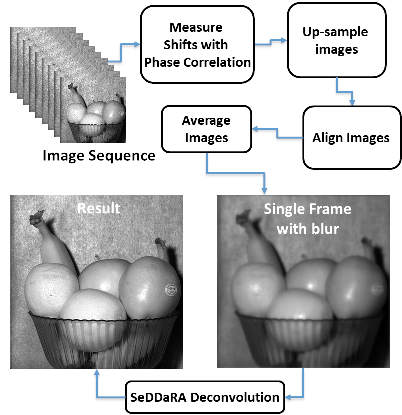}}
\caption{ Graphical depiction of the  ImPASS algorithm.  A sequence of displaced low-resolution images with non-integer translations are up-sampled, aligned, and combined, forming a seemingly blurry single image. Application of SeDDaRA deconvolution produces a high-resolution result. (Image Source: Author)}\label{process}
\end{figure}

Image Phase Alignment Super-sampling (ImPASS) belongs in a class of computational imaging methods that convert a sequence of comparatively low-resolution images into a single high-resolution image.  The low-resolution frames are of static scenes, but vary by sub-pixel translations in the image plane.  These sub-pixel displacements provide information necessary to reconstruct the high-resolution scene.  The processing steps generally consist of aligning, combining, and deconvolution.  ImPASS~\cite{pending} is unique among these approaches in that this algorithm uses Phase Correlation for the alignment step, and SeDDaRA (Self Deconvolving Data Reconstruction Algorithm)~\cite{ao_2002} for the deconvolution step, as depicted in Figure~\ref{process}.  In previous study, application of ImPASS to lab images improved image resolution by a factor of 3.2 whereas resolution subceeded the diffraction limit of the system by a factor up to 2.63.~\cite{aip_2022}  The resolution improvement factor is expressed as the ratio of original image resolution over the processed image resolution.  Applied to microscope image sets, resolution was improved by a factor 2.68 and 1.79 times below the diffraction limit.  These values far exceed performance of similar computational methods that have reported quantitative resolution measurements.~\cite{arxiv_2025}

Image resolution is limited in an optical imaging system  primarily by the diffraction limit as defined by the diameter of the front aperture, and the sampling limit as determined by a pixel size in the imaging plane.  Microscopy methods that capture image resolution beyond the diffraction limit are generally referred to as super-resolution.~\cite{microscopes}  As discussed in reference~\cite{arxiv_2025}, computational methods that produce resolution below the pixel size have had multiple designations including `super-resolution', `geometric super-resolution' and `shift-and-add'.  The term super-sampling~\cite{schuler} is used here to differentiate the computational methods from the super-resolution microscopy terminology.   

The two primary applications for super-sampling are microscopy and remote sensing.  For microscopy, engineering limits put the optical plane resolution limit at around 250~nm.  Customized methods such as STED, STORM,~\cite{jessica} Structured Illumination Microscopy (SIM),~\cite{joby} and Fourier Ptychography~\cite{konda} have been developed to subceed this limit, but require active illumination methods and additional optical elements.  In contrast, super-sampling requires only a translation stage and can be used with any illumination method, providing wide-application and a reduction in costs.  Remote sensing via satellite is hindered as higher resolution requires larger apertures, but larger apertures cannot be accommodated due to weight and size restrictions.  Super-sampling could be employed for remote sensing using a small array of telescopes, or a single scanning telescope.  

Many super-sampling studies have used Phase Correlation image registration (PC) in combination with other deconvolution methods.  For example, in 2005, Young et al~\cite{driggers} used PC coupled with an Error-Energy Reconstruction algorithm.  The survey by Shah et al~\cite{shah} in 2012 described PC in combination with various regularization methods, such as total variation.  In 2025, Brewer~\cite{brewer} used the combination of PC and Lucy-Richardson Deconvolution for application to microscopy.  

In this study, modifications have been made to ImPASS in the hope that greater resolution can be achieved.   Two modifications were made to the registration process such that the sub-pixel location of the correlation peak, which determines the translational differences in two similar images, is determined with greater accuracy. In addition, a change was made to the deconvolution step producing an improved resolution without additional artifacts.  These changes are applied to previous imaging sets to gauge improvement in the process.

\section{Phase Correlation }
Phase correlation~\cite{reddy} is a conventional image registration method with an extended history.~\cite{tong}  Translated images of a static scene are related by 
\begin{equation}\label{feq1}
    I_t(m,n)=I_c(m-\Delta x,n-\Delta y).
\end{equation}
where images $I_c(m,n)$ and $I_t(m,n)$ have a non-integer translational difference of $(\Delta x, \Delta y)$.  According to shift theory,~\cite{weng} application of a Fourier transform to Equation~\ref{feq1} produces
\begin{equation}\label{feq2}
    F_t(u,v)=e^{-i(u \Delta x+ v \Delta y)}F_c(u,v)
\end{equation}
where $(u,v)$ are the coordinates in Fourier space. As such, the translation differences appear as phase changes in Fourier space and  can be derived from the phase difference between the translated images.  The phase component is expressed as 
\begin{equation}\label{feq2a}
    e^{(u \Delta x+ v \Delta y)}=\frac{F_c(u,v)F^*_t(u,v)}{|F_c(u,v)F^*_t(u,v)|}
\end{equation}
where `$*$' represents the complex conjugate.  This difference between the phase components of the translated images appears as a sinusoidal pattern where the frequency and direction of the wave pattern provide information about the alignment difference.  Application of a second Fourier transform
to the phase difference produces an array with a  correlation peak whose coordinates are $(\Delta x, \Delta y)$.  Since this array is provided in a discreet array of pixels, the peak will fall onto several neighboring pixels.  As such, the accuracy of this image registration is dependent upon determining the sub-pixel location of this peak accurately.  There have been many papers devoted to this endeavor.  In 2002, Foroosh~\cite{foorosh} provided an analytical model for subpixel registration and to produce a discreet estimate of the peak location.  In 2003, Takita~\cite{takita}  applied a two-dimensional fitting function to the peak to more accurately determine the peak location. In 2008,  Guizar-Sicairos evaluated three versions of improving accuracy by up-sampling the correlation peak.  In 2016, Ren~\cite{ren} evaluated several methods concluding that Gaussian-based dual-side interpolation best fit their experiments.  A review of methods is provided by HajiRassouliha~\cite{amir}.  A comparison of these methods is beyond the scope of this paper, however an earlier comparison of methods was presented by Reed in 2010.~\cite{reed}

In contrast to these more sophisticated methods, I have had good success locating the sub-pixel location using the computationally-efficient centroid function.~\cite{shortis} After determining the maximum-valued pixel in the correlation peak array $I_{cp}(m,n)$, the subpixel location is found using
\begin{equation}\label{feq5}
    \Delta x=\frac{\sum_{m=m_p-a}^{m_p+a}m \; I_{cp}^p(m,n)}{\sum_{m=m_p-a}^{m_p+a} I_{cp}^p(m,n)}  \: \textrm{and} \: \Delta y=\frac{\sum_{n=n_p-a}^{n_p+a} n \; I_{cp}^p(m,n)}{\sum_{n=y_p-a}^{n_p+a} I_{cp}^p(m,n)}
\end{equation}
where the value $a$ provides the size of the local search region, typically set to $a=2$.~\cite{shannon}  When $p=1$, this equation is the standard centroid. For $p=2$, this equation is referred to as the centroid of squares, but non-integer values are also possible, as discussed in Section~4.1.

Comparisons of different registration methods, and variations of those methods, are beyond the scope of this project.  Here the objective is to determine if an improvement in alignment accuracy results in increased resolution.  This helps determine practical limitations of ImPASS.  If the improvement in accuracy alignment does not lead to improvement in resolution, then the primary limitation may lie with the physical characteristics of the system, such as noise or optical aberrations.

\section{Error Reduction Model}
\begin{figure}[htbp]
\center{\includegraphics[width=19pc]{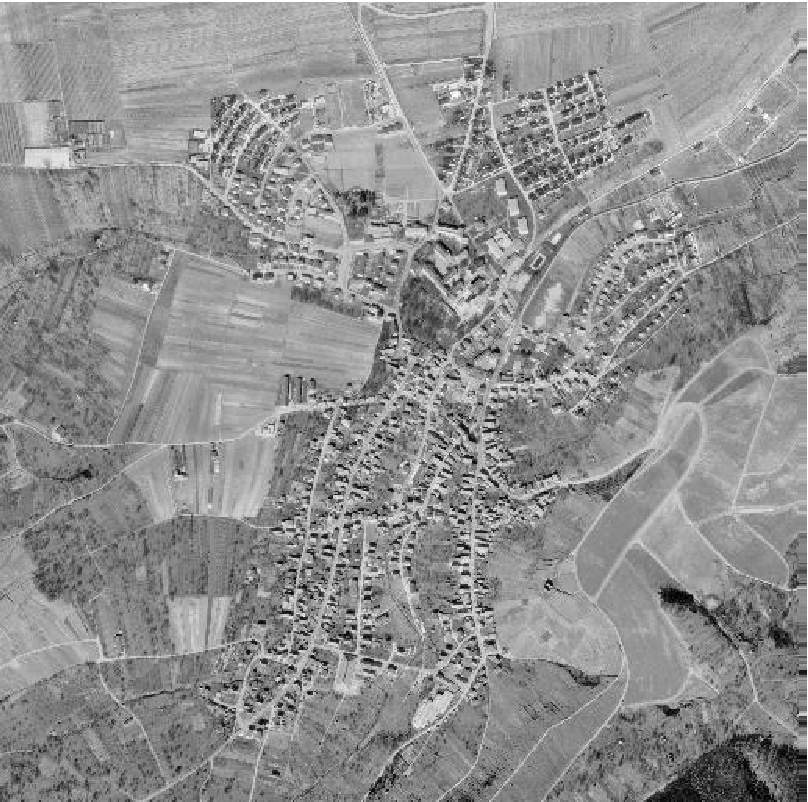} \hspace{1mm} \includegraphics[width=19pc]{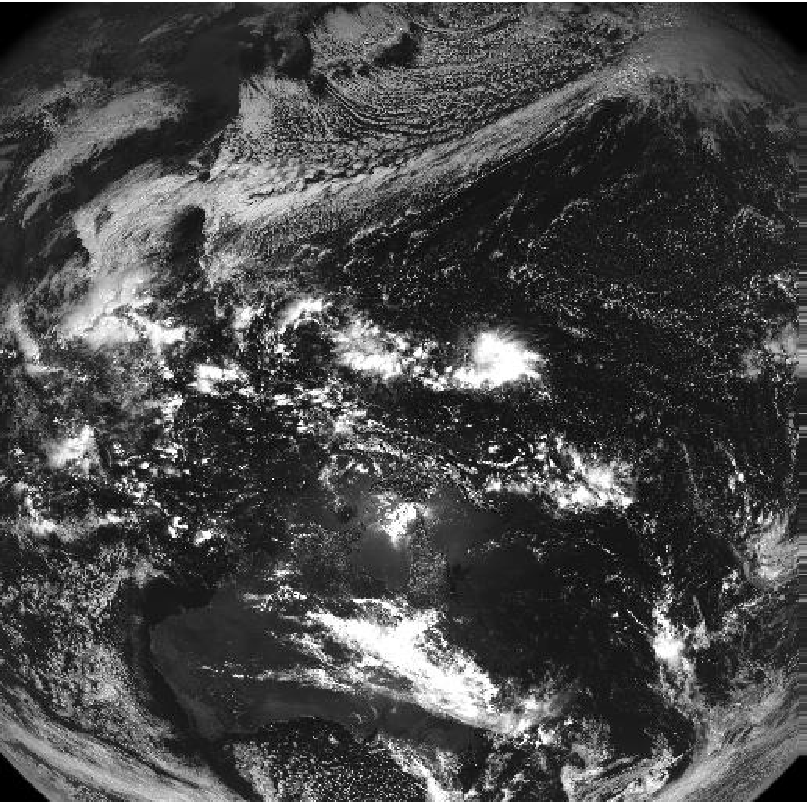}}
\caption{ A single frame of the 16X down-sampled sequences from the Kernen set (left) and the Himawari set (right). (Left Image Source: From https://www.leo-bw.de/kartenvergleich and available through the Creative Commons Attribution 3.0 Unported license.)(Right Image Source: From Japan Meteorological Agency and is in Public Domain)}\label{kernen}
\end{figure}

Simulated data sets were created to test improvements in the image registration process.  The process starts with a single truth image having dimensions greater than $8208^2$ pixels per side,  where $X^2$ is shorthand for $X \times X$ pixels. The truth images consisted of an aerial view of with the name `Kernen stetten 1968 ortho.jpg' from Wikipedia of size $8639^2$ pixels, nicknamed 'Kernen', and a view of Earth from the Himawari weather satellite of size 11000$^2$ pixels.  A 64-frame sequence of images with size $512^2$ pixels was created with the following steps:
\begin{enumerate}
  \item Original image is shifted a single pixel in the $\hat{x}$ and $\hat{y}$ dimensions.
  \item The image is center cropped to size $8192^2$ pixels.
  \item The image is down-sampled by a factor of 2 in each direction four times to reach an image size of $512^2$.
  \item The image is saved as a single frame in the sequence.
  \item The process repeats until 64 frames are created.
\end{enumerate}
As such, the sets have a down-sampled factor of 16 with pixel shifts between frames being exactly 1/16th of a pixel.  Single frames of the down-sized image sets are shown in Figure~\ref{kernen}.
\section{Algorithm Improvements}
Several modifications have been made to this algorithm since the previous ImPASS publication.~\cite{arxiv_2025}  

\subsection{Centroid Power}
\begin{figure}[htbp]
\center{\includegraphics[width=28pc]{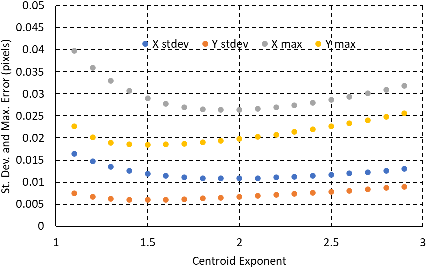}}
\caption{ Total displacement error for the Kernen set as a function of the centroid power. The data suggests that a value of $p=2$ will improve accuracy for Phase Correlation.}\label{power}
\end{figure}

In the previous version, ImPASS used the standard centroid function to locate the sub-pixel location of the centroid peak.  To optimize this task, the value of $p$ in Equation~\ref{feq5}  was varied from 1 to 2.9 while applying the phase correlation repeatedly to the two test sets. The measured displacements were compared to ideal values using the standard deviations and maximum errors as metrics.  The Kernen results are shown in Figure~\ref{power} for the $\hat{x}$ and $\hat{y}$ axes as a function of the centroid power $p$.  The plot shows that the error is minimized for values near $p=1.5$ for shifts in the $\hat{y}$ direction and near $p=2$ for $\hat{x}$.  When applied to the Himawari set, the minimal points were  between 1.8 and 2.2, making this optimization dependent on the scene and orientation. The change from $p=1$ to $p=2$ in this case produced a significant 50\% reduction in error in the $\hat{x}$ dimension and a 17\% reduction in the $\hat{y}$ dimension.   As such, $p=2$, the centroid of squares, is the new standard for this algorithm. 

\subsection{Multi-pass Phase Correlation}
\begin{figure}[htbp]
\center{\includegraphics[width=16pc]{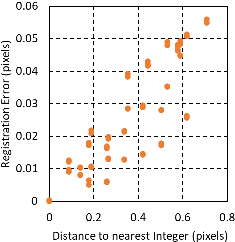} \hspace{1mm} \includegraphics[width=16pc]{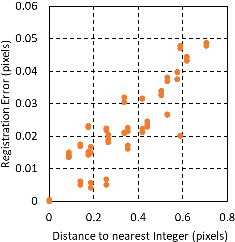}}
\caption{ Total displacement error  as a function of distance to integer pixel for the Kernen set (left) and the Himawari set (right). The plots show that alignment errors increase as the fractional shift increases.  }\label{displace}
\end{figure}

The second modification resulted from varying the designated control image of a sequence.  In a typical application of Phase Correlation image registration, one frame of the image sequence is chosen as the control image with translation differences of other frames being measured in reference to that control image.  As has been shown~\cite{tong, zden},  the registration accuracy depends on the fractional shift, where smaller shifts produce increases accuracy.  For example, shifts that are off by 0.1~pixels can be registered with greater accuracy than when shifts are off by 0.5~pixels.  To demonstrate, a single application of Phase Correlation was applied to the simulated data sets with displacements recorded as $x_d$ and $y_d$.  The difference between measured and true values, $x_t$ and $y_t$, were recorded as $\Delta x_d = x_t-x_d$ and $\Delta y_d = y_t-y_d$.  The total error for each registration is calculated using  $e_d = \sqrt{\Delta x_d^2 + \Delta y_d^2}$.  Figure~\ref{displace} shows the alignment error for the Kernen (left) and Himawari (right) sets as a function of the fractional distance from an integer pixel shift, $d_d = \sqrt{(\lfloor x_t \rceil - x_t)^2 + (\lfloor y_t \rceil - y_t)^2}$ where $\lfloor \nu \rceil$ rounds the value $\nu$ to the nearest integer.  A linear relationship is apparent for each set.  The Kernen set produces a slope of  0.073~error/distance while the Himawari set presents a slope of 0.058~error/distance, suggesting that registration accuracy can be scene-dependent.

For multi-pass application of phase correlation, each frame takes a turn as the control frame.  Given a sequence of $Q$ images, there are $Q$ applications of Phase Correlation producing $Q$ sets of displacements.  These values can be assembled into an array $D(c, z)$ for each image axis where $z$ is the frame index and $c$ is the index of the control image.   This data is shifted to align the sets to a common control image, such as $z=0$.  For each $c$, I created an array $D_{new}(c,z) = D(c,z) - D(c,0)$.  The average is taken using
\begin{equation}\label{eq1}
  D_{fin}(z) = 1/Q \sum_{c=0}^{Q-1} (D(c,z) - D(c,0))
\end{equation}
to produce the final translation corrections.

\begin{figure}[htbp]
\center{\includegraphics[width=16pc]{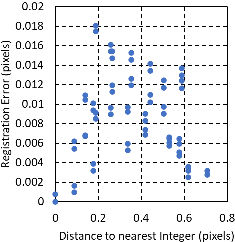} \hspace{1mm} \includegraphics[width=16pc]{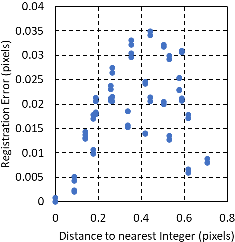}}
\caption{ Total displacement error for the Kernen set (left) and the Himawari set (right). The additional processing removes the linearity and reduces alignment error.}\label{displace1}
\end{figure}

The total errors derived from $D_{fin}(z)$ are displayed in graphs in Figure~\ref{displace1}, revealing that the linear dependence has been removed.  The maximum error has been reduced by a factor of 3.11~for the Kernen set and 1.44~for the Himawari set, producing accuracy to 1/56th and 1/29th of a pixel, respectively.   The average error diminished by a factor of 3.02~for the Kernen set and~1.26 for the Himawari set.

\subsection{Deconvolution Noise Threshold}
The third modification involves a change to the deconvolution method.  After SeDDaRA extracts a point spread function $D(u,v)$, the processed image is calculated in frequency space using the pseudo-inverse function (also referred to as the Wiener filter) using
\begin{equation}\label{eq4}
F(u,v) = \frac{G(u,v) \; D^*(u,v)}{|D(u,v)|^2+C_2 D_{ave}}
\end{equation}
where $(u,v)$ are the coordinates in Fourier space, $G(u,v)$ is the combined image and $D_{ave}$ is the average of the absolute value of $D(u,v)$.  The parameter $C_2$ prevents the over-amplification of noise in the areas of the image where the signal level is below the noise, typically set by trial error to be between 0.001 and 0.1.  Instead of conventionally adding $C_2 D_{ave}$ in the denominator, this value is used instead as a threshold.  Any values in $|D(u,v)|^2$ that are lower than $C_2 D_{ave}$ are replaced by $C_2 D_{ave}$.  The idea is that this change will prevent noise from below the threshold level from adversely affecting the processing,  while areas above the threshold are deconvolved by ideal inverse filter.   

\begin{figure}[htbp]
\center{\includegraphics[width=16pc]{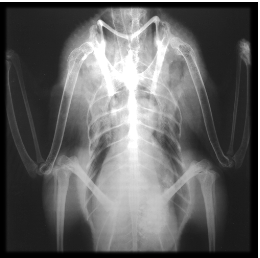} \hspace{1mm} \includegraphics[width=16pc]{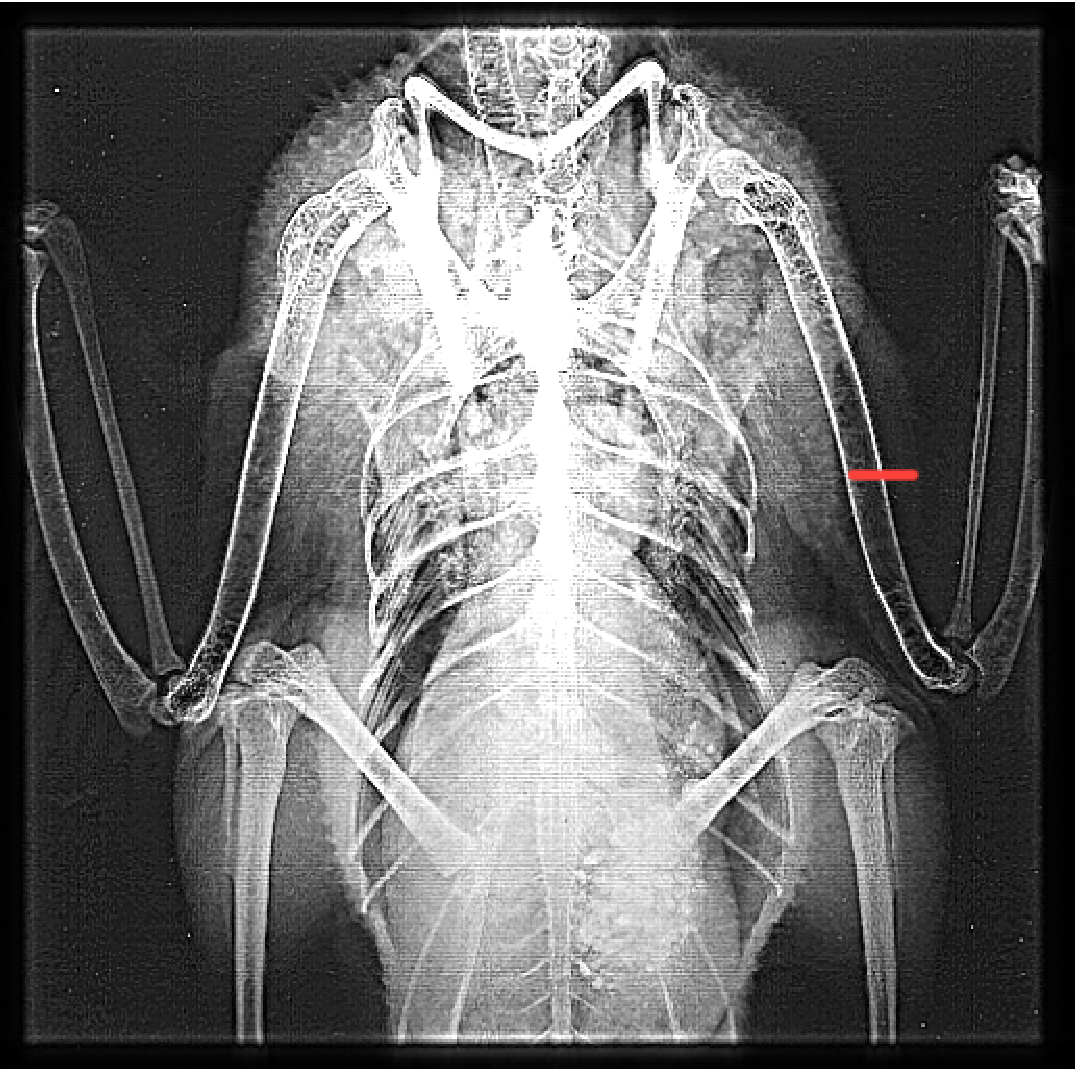}}
\caption{(Left) A digitized x-ray of a Muscovy duck.  (Right) A SeDDaRA deconvolution of the image.  The red segment indicates the location where qualitative resolution was measured. (Source: Example image from NIH Image)}\label{duck}
\end{figure}

Figure~\ref{duck} shows an x-ray of a Muscovy duck~\cite{ao_2002} and the SeDDaRA deconvolution.  When the deconvolution was performed with the original method, the full-width half maximum (FWHM) of the profile plot, represented by the red line, was 4.19 pixels.  With the new approach, the FWHM was 3.63 pixels, improved by a factor of 1.15, with no apparent additional artifacts as a result.  

\section{Application to Laboratory Image Sets}\label{thor}
The modified ImPASS algorithm was applied to sets of laboratory images created in 2023 for a spectral study, where the target was a USAF resolution chart placed on an XY translation stage. A machine vision camera with a 50~mm lens captured each frame as the chart was moved in discreet steps around the object plane.  The lens had a 650~nm filter with a bandwidth of 35.5~nm.    The camera had a 1920 X 1028 pixel focal plane array where pixel size was 5.86$^2$ micrometers in the image plane producing an angular resolution of 72.3~microradians.  Each set consisted of 64 displaced images where the total range of translation was 0.28~mm in the horizontal ($\hat{x}$) dimension and 0.27~mm in the vertical ($\hat{y}$) direction.

The iris of the aperture was changed for each set, allowing the results to be compared to the diffraction limit of the camera for a range of f-numbers.  As the object was not placed at infinity, the effective f-number was calculated from 
\begin{equation}\label{eq2}
  f\#_{eff} = (1+|M|)f\#
\end{equation}
where magnification $M$ was 0.0560.  

The Abbe diffraction limit~\cite{abbe} is given by
\begin{equation}\label{eq3}
  \theta = \lambda f\#_{eff}/f  
\end{equation}
where $f$ is the focal length and $\lambda$ is light wavelength.

The original images were cropped to $540^2$ pixels for easier processing, aligned using the new phase correlation modifications, up-sampled  by a factor of $M_{ss} = 8$ to  $4320^2$ pixels, and combined using averaging.  SeDDaRA deconvolution was then applied to the combined image using processing parameter values of $RofI=28$ pixels and $C_2=0.008$.  The radius of influence $RofI$ limits the influence of noise in the extracted point spread function by setting a threshold whereby all values below the threshold are set to 0.  The parameter $C_2$ appears in the denominator of the pseudo-inverse filter to prevent noise amplification as shown in Equation~\ref{eq4}.

\begin{figure}[htbp]
\center{\includegraphics[width=16pc]{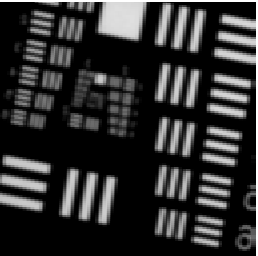} \hspace{1mm} \includegraphics[width=16pc]{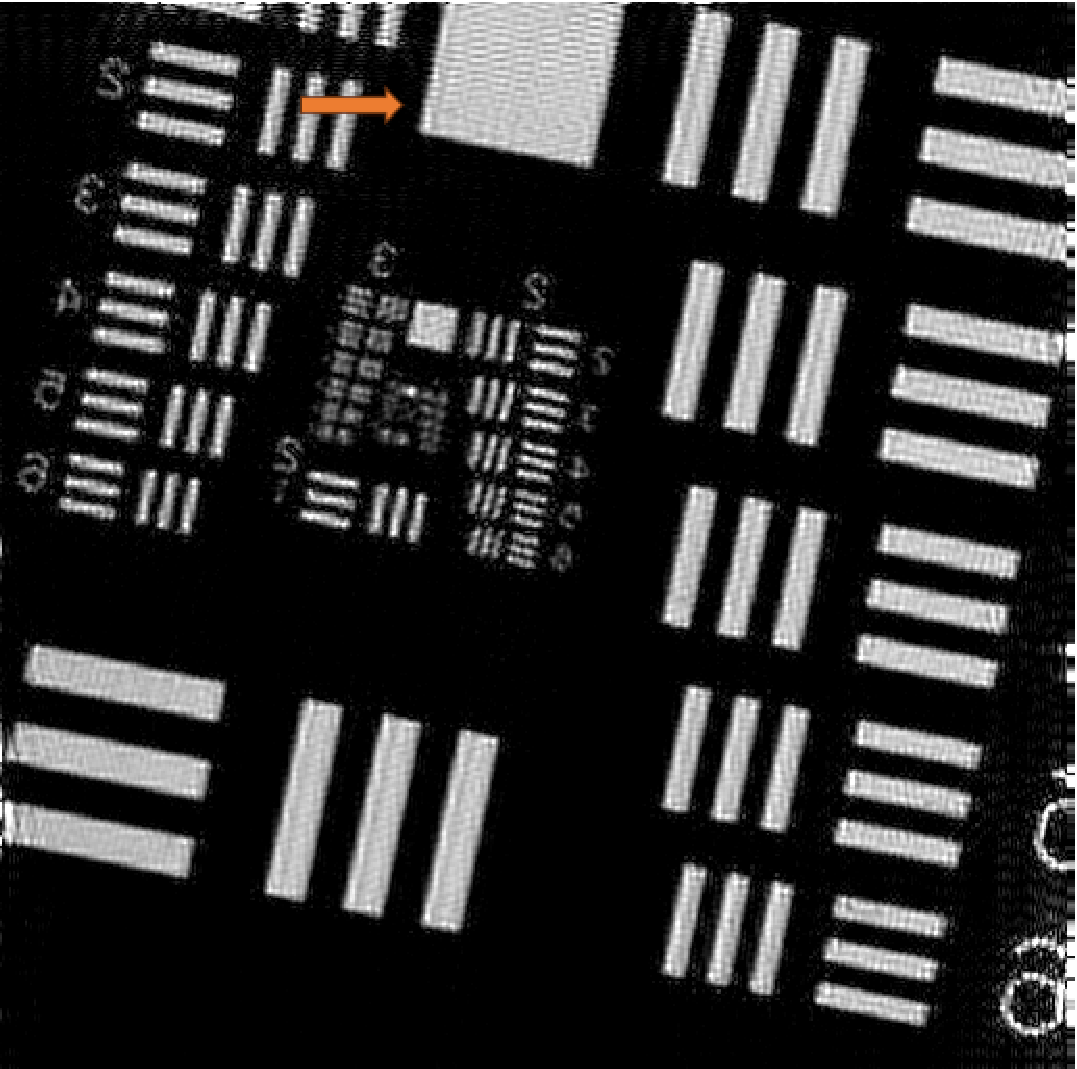}}
\caption{ Left image is a 153X175 pixel region of an original frame from the $f\#=6$ laboratory image set.  Right image is the similar region of the ImPASS processed image, showing much improvement in the resolution. }\label{visual}
\end{figure}
\begin{figure}[htbp]
\center{\includegraphics[width=20pc]{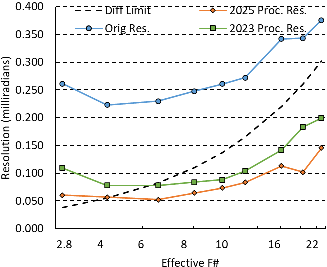}}
\caption{Comparison of the previous and modified processed image resolution to the original image resolution and Abbe's diffraction limit as a function of effective f-number.  The modified processing improved final resolution between 20\% and 81\% as compared to the previous iteration.}\label{resgraph}
\end{figure}

Figure~\ref{visual}(left) shows a cropped version of an original image, revealing pixelization, from the laboratory set with f-number of 6.   The same region of the processed version, on right, shows significant improvement in resolution. The bars in series 2 are fully resolved as are the horizontal bars of series 3, group 1.   A processed image is produced for each f-number setting.    Resolution is measured using the slant-edge technique described in Reference~\cite{aip_2022}.  The graphic arrow shows the location where the resolution measurement was taken. The results are plotted in Figure~\ref{resgraph} alongside resolution of the unprocessed images, the diffraction limit for this system, and the processed resolution acquired before modifications to the process. The modifications produce significant resolution gain for all f-numbers.

\begin{table}[htbp]
  \centering
  \begin{tabular}{|l|c|c|c|c|c|c|c|}
  \hline
 $f^{\#}_{eff}$ & \thead{Diff. Limit} & \thead{Orig. Res.} & \thead{Prev. Proc. \\ Res.} & \thead{Mod. Proc. \\ Res.} & \thead{Orig./ Mod. \\ Proc.} & \thead{ Diff. Lim./ \\ Mod.  Proc.} & \thead{Prev. Proc./ \\ Mod. Proc.} \\ \hline 
  2.96 & 0.038 & 0.261 & 0.109 & 0.0605 & 4.32 &  0.635 & 1.81\\
  4.22 & 0.055 & 0.223 & 0.078 & 0.0560 & 3.95 & 0.972 & 1.38\\
  6.34 & 0.082 & 0.230 & 0.078 & 0.0521 & 4.41 & 1.582 &  1.49\\
  8.45 & 0.110 & 0.248 & 0.084 & 0.0645 & 3.84 & 1.703 & 1.30\\
  10.56 & 0.137 & 0.261 & 0.088 & 0.0735 & 3.55 & 1.868  & 1.20\\
  12.67 & 0.165 & 0.272 & 0.104 & 0.0832 & 3.27 & 1.979 & 1.25\\
  16.90 & 0.220 & 0.342 & 0.142 & 0.1129 & 3.03 & 1.946 & 1.26\\
  20.06 & 0.261 & 0.343 & 0.183 & 0.1016 & 3.38 & 2.568 & 1.81\\
  23.23 & 0.302 & 0.376 & 0.199 & 0.1454 & 2.58 & 2.077 & 1.37\\
  \hline
\end{tabular}

  \caption{Resolution values as a function of effective f-number. Columns 2 to 5 have resolutions in microradians for the diffraction limit, unprocessed, processed using previous iteration, and processed with modified algorithm. Columns 6 to 8 are ratios of the various resolutions. }\label{restable}
\end{table}

As shown in Table~\ref{restable}, overall improvement of resolution ranged from 2.58 to 4.41.  The f-number 19 set subceeded the diffraction limit by a factor of 2.568.  When compared to resolution measurements taken before the modifications (2023), resolution was improved between 20\% and 81\% producing an average improvement of 43\% and median value of 37\%.

\section{Application to Microscopy Image Set}
In reference~\cite{arxiv_2025}, ImPASS was applied to a sequence of images of a resolution chart as captured by an Olympus Inverted Fluorescence Microscope IX-83 with an Olympus UPlanFL 4x lens having a  numerical aperture of 0.13.  The Hammatsu ORCA-Flash 4.0 v3 camera had a focal plane array consisting of 2034 X 2042 pixels that measured 6.5 $\mu$m on each side.  The optical diffraction limit of the object plane was 2.50 microns. Using an XY translation stage, 80 slightly-displaced frames were created with diagonal steps of 990~nm forming an `X' pattern.  These frames were cropped to 532$^2$ pixels and up-sampled by a factor of $M_{SS} = 8$.  The unprocessed frames produced a resolution of 3.73 microns.  The processed result had a resolution of 1.39 microns, well below the diffraction limit.

\begin{figure}[htbp]
\center{\includegraphics[width=16pc]{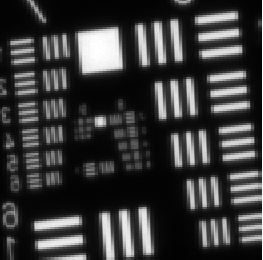} \hspace{1mm} \includegraphics[width=16pc]{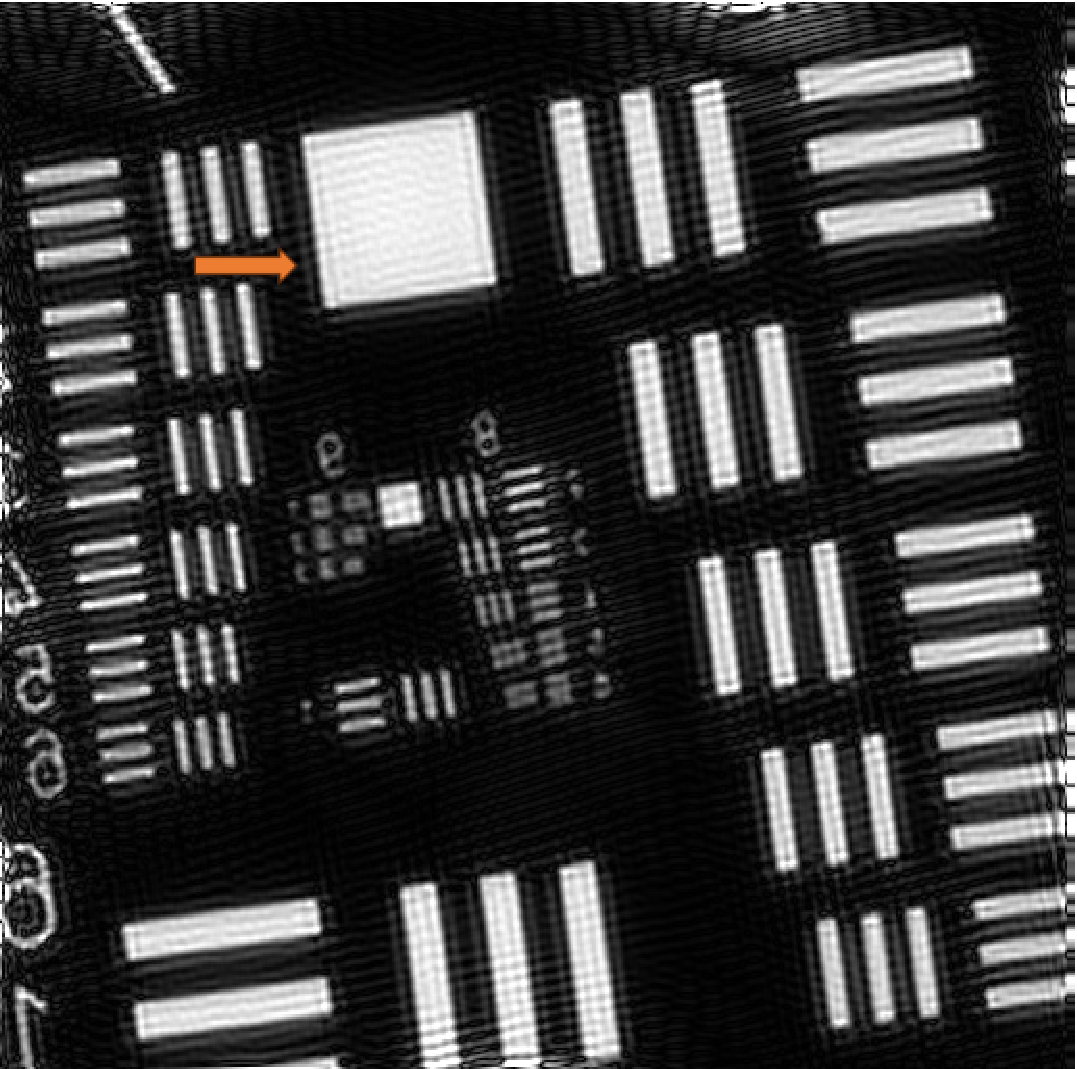}}
\caption{ Application of modified ImPASS to microscope set.  The left image is a cropped 128$^2$ pixel region of an original frame showing the central region.  The right image is a cropped 1024$^2$ pixel image showing the same region of the processed image.  The graphic area points to the approximate location where resolution was measured.}\label{mic}
\end{figure}

The modified algorithm was applied to the set with the result shown in Figure~\ref{mic}.  Application of the slant-edge resolution measurement, marked by an arrow in Figure~\ref{mic}(right), found a resolution of 4.794 pixels or 0.974 microns.  This is an improvement of 3.83 times compared to the original, subceeding the diffraction limit by a factor of 2.571.  The modified ImPASS version produced 74\% greater resolution than the previous iteration.

\section{Application to MicroFilm Image Set}
Application of ImPASS is not limited to resolution charts.  Targets will have different distribution of spatial frequencies requiring a change of reference image and SeDDaRA parameters.  The SeDDaRA reference frame provides a model frequency distribution with pixel-level resolution at the up-sampled image pixel size. Choosing a reference frame, given limited study of this method, is accomplished by trial-and-error.  

\begin{figure}[htbp]
\center{\includegraphics[width=16pc]{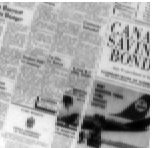} \hspace{1mm} \includegraphics[width=16pc]{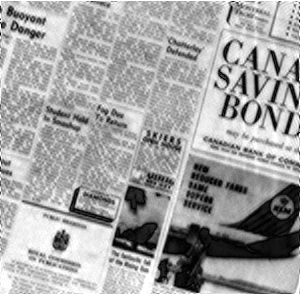}}
\caption{ Application of modified ImPASS to a strip of microfilm positioned on an XY translation stage.  The left image is an original frame, cropped to a 128$^2$ pixel central region.  The right image is a processed 1024$^2$ pixel image showing the same region.  The `KLM' logo can be identified in the processed image as well as words with larger fonts.}\label{fish}
\end{figure}

For demonstration, the resolution chart was replaced with a microfilm strip in the setup described in Section~\ref{thor}.  The microfilm has images of a Canadian newspaper, possibly from World War II.  Thirty-two images were captured as the XY translator was moved step-by-step in a diagonal pattern.  The average step size was 0.172 pixels with a standard deviation of 0.266 pixels.   Figure~\ref{fish}(left) shows the center portion of an original frame.  The images were cropped to 532$^2$ pixels, up-sampled by a factor of $M_{ss}=8$, aligned, and combined.   Using a digital image of a newspaper as the reference image, SeDDaRA was applied to the combined image.  The reference image was rotated 10.5~degrees to match the rotation of the microfilm to reduce artifacts.    

As shown in Figure~\ref{fish}(right), the processed image is significantly improved.  The `KLM' can be read on the tail fin of the airplane, the words "Canadian Bank of" and "may be purchased" can be read in the advertisement.  For smaller fonts, groups of letters can be identified, but the specific words cannot. 

\section{Comparison to Similar Methods}   
While there are a large variety of research articles on super-sampling, the majority have used qualitative measures like Mean-square error (MSE), Peak signal-to-noise ratio (PSNR), and Structured Similarity Indexing Method (SSIM) to gauge improvement.~\cite{umme}  These comparative metrics do not measure image resolution directly.  Of the papers that have, Stankevich~\cite{stan} found a 1.49 times resolution improvement, while Brewer's approach~\cite{brewer} measured a 1.40 times improvement.  The ImPASS results significantly exceed these values, producing a 3.83 times improvement for the microscope image, and up to 4.41 times improvement for the laboratory experiment.

ImPASS also compares favorably for application of Super-resolution microscopy (SRM).  Less-expensive methods like Pychtography and SIM are fundamentally limited to achieving a resolution that subceeds the diffraction limit by a factor of two.~\cite{arxiv_2025}  With the modifications, ImPASS achieved a resolution that was a factor of 2.57 times below the diffraction limit, using a 4X objective lens. A logical next step is to test this algorithm on a system using a 40X objective or similar to determine whether processed image resolution can subceed the 250~nm resolution barrier.  

ImPASS should also be considered a viable alternative to laser-based methods such as STED and STORM.  While these methods can obtain resolutions that subceed the diffraction limit by a factor of 5, these methods can be orientation-dependent, have variable applicability, may photobleach the sample, and expensive.  The choice of illumination wavelength is also limited.  In contrast, ImPASS can be used with any illumination wavelength and in combination with other imaging methods such as confocal microscopy.

\section{Conclusions}
Three modifications have been made to the ImPASS algorithm since the 2025 paper.  Two modifications were made to the registration method.  When applied to simulated sets with known displacements, alignment accuracy improved by 1.44 for one set and 3.11 for the second set.  The third modification changed how the denominator of the pseudo-inverse filter is applied for SeDDaRA.   In a single application of SeDDaRA, this improved resolution by a factor of 1.15.  

The modified ImPASS algorithm was applied to previous image sets to gauge improvement.  Applied to the laboratory image sets with varied f-numbers, the modified version improved resolution by a factor that ranged from 2.58 to 4.41 as compared to the original image resolution, and subceeded the diffraction limit by a factor up to 2.568.  The modified algorithm improved resolution by a factor that ranged from 1.20 to 1.81 when compared to previous results.  Applied to the microscopy image set from  article~\cite{arxiv_2025}, application of modified ImPASS produced a resolution improvement of 3.83 while subceeding the diffraction limit by a factor of 2.571.  The modified version improved resolution by a factor of 1.74 over the previous algorithm.

These improvements resulted from modifications to the registration process and the deconvolution method.  This demonstrates that previous results were not limited by image noise or the physical optical setup.  There is no evidence that these factors are limiting the current results.  Improved super-sampling resolution can be achieved with improved image registration.  The phase correlation method used here achieved accuracy up to 1/56th of a pixel.  Greater accuracy may be obtained using active measuring of this quantity, at least in the case of microscopy where an interferometer could measure displacements.

\section*{Acknowledgments}
The author appreciates the contributions from  Dr. Giuliano Scarcelli and Dr. Jiarui Li with the Fischell Department of Bioengineering, University of Maryland, College Park, MD, USA for donating the microscopy image sets used in this study.


\end{document}